\pdfoutput=1 
\documentclass{jfm}
\usepackage{graphicx}
\usepackage{epstopdf, epsfig}
\usepackage{amsmath}
\usepackage{bm}
\usepackage{color}

\def \be {\begin{equation}}
\def \ee {\end{equation}}
\def \ba {\begin{array}}
\def \ea {\end{array}}
\def \bea {\begin{eqnarray}}
\def \eea {\end{eqnarray}}
\def \bse {\begin{subequations}}
\def \ese {\end{subequations}}
\def \bsea{\begin{subeqnarray}} 
\def \esea{\end{subeqnarray}} 

\def \bcM{\boldsymbol{\mathcal M}}
\def \bPsi{\boldsymbol{\Psi}}
\def \bE{\boldsymbol{E}}

\shorttitle{Inertial wave super-attractor}
\shortauthor{B. Favier, S. Le Diz\`es}

\title{Inertial wave super-attractor\\ in a truncated elliptical  cone}

\author{Benjamin Favier\aff{1}\corresp{\email{benjamin.favier@cnrs.fr}} and St\'ephane Le Diz\`es\aff{1}}
\affiliation{\aff{1} Aix Marseille Univ, CNRS, Centrale Marseille, IRPHE, Marseille, France}

\begin{document}

\maketitle

\begin{abstract}
We consider inertial waves propagating in a fluid contained in a non-axisymmetric three-dimensional rotating cavity.
We focus on the particular case of a fluid enclosed inside a truncated cone or frustum, which is the volume that lies between two horizontal parallel planes cutting an upright cone.
While this geometry has been studied in the past, we generalise it by breaking its axisymmetry and consider the case of a truncated elliptical cone for which the horizontal sections are elliptical instead of circular.
The problem is first tackled using ray tracing where local wave packets are geometrically propagated and reflected within the closed volume without attenuation.
We complement these results with a local asymptotic analysis and numerical simulations of the original linear viscous problem.
We show that the attractors, well-known in two dimensional or axisymmetric domains, can be trapped in a particular plane in three-dimension provided that the axisymmetry of the domain is broken.
Contrary to previous examples of attractors in three-dimensional domains, all rays converge towards the same limit cycle regardless of initial conditions, and it is localised in the bulk of the fluid.
\end{abstract}

\section{Introduction}

Inertial waves and internal gravity waves are waves propagating in rotating and stratified fluids, respectively.
In three dimensions, when forced locally in a uniformly rotating (or stratified) fluid, these waves have the particularity to propagate along a double cone that makes a constant angle with respect to the axis of rotation (or to the direction of stratification).
This property allows to show that, when propagating within a closed container, wave packets may converge after multiple reflections on solid boundaries to a particular surface called attractor in 2D \citep{Maas1995} or axisymmetric geometries such as the spherical shell \citep{Rieutord1997}.
One of the simplest geometries giving rise to attractors is a rectangular container with a sloping boundary.
This trapezoidal geometry has been the subject of numerous works since the first study in a stratified fluid by \cite{Maas1997}.
The theoretical attractors obtained by ray tracing have been observed both experimentally \citep{Hazewinkel2008,Hazewinkel2010,Scolan2013} and numerically \citep{Drijfhout2007,Grisouard2008}.
Similar results were also obtained in rotating fluids in the same geometry \citep{maas2001wave,Manders2003,Manders2004} or in its axisymmetric version
\citep{Klein2014,sibgatullin2019influence,Boury2021,pacary2023}.
Attractors have also been found to be generic features of inertial waves in spherical shells \citep{Rieutord1997,Rieutord2000}.
Early studies concerned attractors localised close to the equator \citep{Stern1963,Bretherton1964,Stewartson72} 
which possess similar properties as 2D attractors \citep{Rieutord2002}.
But the picture seems more complex in a spherical shell
owing to the presence of the rotation axis in the domain and of a critical latitude singularity issued from the inner sphere \citep{Rieutord2010,Rieutord2018,He2022,He2023}. 
The robustness of attractors has been analysed with respect to wall friction \citep{Beckebanze2018}, nonlinearity \citep{Grisouard2008,jouve2014direct,Favier14,Beckebanze2021,Ryazanov2021}
and instabilities \citep{Brouzet2016,Brouzet2017,Dauxois2018}. 
The 2D framework has also been used to obtain most of the available mathematical results \citep{Maas1995,manders2003wave,Maas2009,bajars2013appearance,Beckebanze2016,ColinDeVerdiere2020,Dyatlov2022,Makridin2023}.

While fundamental in nature, studies about attractors are motivated by their ability to focus energy at small length scales, which could potentially impact the dissipative properties of many geophysical systems.
Internal gravity waves and attractors can break and lead to turbulence \citep{staquet2002internal,Brouzet_2016}, and as such are known to play an important role in the energy budget of the ocean where they are often excited by tides \citep{Wunsch75}.
Attractors can for example be excited by tidal waves in a paraboloidal basin \citep{maas2005wave} or between two ridges \citep{Echeverri2011}.
Their occurrence has been analysed in the configurations of the Mozambique Channel \citep{Manders2004b} and of the Luzon strait \citep{Tang2010,Wang2015}. 
In the astrophysical context, inertial waves and attractors could be important in the synchronisation processes of rapidly-rotating astrophysical objects as they provide a way to rapidly dissipate energy \citep{Zahn1975,Ogilvie04}.

Most of the works mentioned above have considered a 2D or a 3D framework with symmetry (axisymmetric or invariant along one direction).
In that case, rays were assumed to remain confined within a particular 2D plane since the system is effectively invariant along the third direction.
As soon as the rays are assumed to propagate out of this 2D plane, it becomes important to consider the 3D reflection law of localised wave beams \citep{Manders2004,maas2005wave}.
This was done in an axisymmetric geometry by \cite{maas2005wave} and \cite{Rabitti2013,Rabitti2014}, and in a 3D rectangular domain with a sloping wall but invariant along one direction by \cite{Manders2004} and \cite{Pillet2018}.
For both the axisymmetric spherical shell and the 3D trapezoidal basin that has a uniform shape in a transverse direction, the authors showed that the attractor is possibly trapped in a specific plane which depends on the initial conditions of the ray tracing protocol.
To our knowledge, the only study who considered a completely 3D geometry is the recent work of \cite{Pillet2019}.
They considered a trapezoidal geometry, but for which the sloping plane is also inclined in the transverse direction, thus breaking the symmetry on which previous studies were implicitly constructed.
In that case, the ray beams tend to drift along the twice-inclined boundary towards the vertical boundary eventually closing the domain.
The rays accumulate there around a trapped attractor close to the vertical boundary \citep{Pillet2019}, which is not due to a global focusing but instead due to the arrest by the vertical side boundary of the continuous shift of the wave beams.

In the present work, focusing on the purely rotating case with inertial waves only, we consider a 3D non-axisymmetric geometry and show that we obtain a global focusing, regardless of initial conditions, of the wave beams along a local curve (as opposed to a surface in the axisymmetric case).
Contrary to the work of \cite{Pillet2019}, our attractor does not rely on the confinement induced by a vertical boundary but is trapped in the bulk of the fluid domain.
We thereby provide the first evidence of a super-attractor, that is a 1D curve on which all rays tend to focus after multiple reflections on solid boundaries, regardless of their initial positions and orientations within the fluid volume. 
In \S \ref{sec:formulation}, we present the framework, the 3D reflection law and our geometry (a truncated elliptic cone).
In \S \ref{sec:tracing}, we demonstrate that, in our geometry, ray beams corresponding to waves of a given frequency do converge for almost all initial conditions to a unique limit cycle. This limit cycle only depends on the geometry and the frequency of the waves.
An analytic formula for the contracting factors obtained in the two focusing directions is derived in this section
and compared to numerical ray tracing results.
In \S \ref{sec:dns}, numerical simulations of the linear viscous problem are presented.
We show that the viscous response to a global forcing tends to be focused onto the super-attractor.
A preliminary scaling with respect to the Ekman number is also proposed for the velocity amplitude, before we finally present our conclusions.

\section{Formulation of the problem and methods}
\label{sec:formulation}

In this section, we describe the equations of motion, the reflection law of a localised wave packet and the particular geometry considered for the rest of the study.

\subsection{General problem and equations}

We consider the incompressible flow of a fluid of constant kinematic viscosity $\nu$ contained inside a closed container rotating at a constant rate $\bm{\Omega}=\Omega\bm{e}_z$.
While we expect our results to remain valid when the fluid is stratified in density, due to the similarity in the dispersion relations and propagation properties, we focus on the purely rotating case here for simplicity.
Using $1/(2\Omega)$ as the time scale and the characteristic length scale $a$ of the container as the reference length scale, and focusing on infinitesimal perturbations to the solid body rotation flow, the linearised dimensionless equations of motion in the rotating frame are
\begin{align}
\label{eq:momentum}
\frac{\partial\bm{u}}{\partial t}+\bm{e}_z\times\bm{u} & =-\nabla P+E\nabla^2\bm{u} \\
\label{eq:mass}
\nabla\cdot\bm{u} & =0
\end{align}
where $\bm{u}$ is the perturbation velocity, $P$ is the pressure incorporating the centrifugal acceleration and $E=\nu/(2\Omega a^2)$ is the Ekman number.
We shall focus on the response to an harmonic boundary forcing of dimensionless frequency $\omega$ and for small Ekman numbers.
The frequency of inertial waves is bounded by twice the rotation rate and we therefore focus on dimensionless frequencies comprised between 0 and 1.

Information on the solution can be obtained in an inviscid framework by monitoring the propagation of localised wave beams, as detailed in the next section. 

\subsection{Reflection law of a localised beam}

Our description mainly follows the approach of \cite{Manders2004}, \cite{maas2005wave}, \cite{Rabitti2014} and \cite{Pillet2018,Pillet2019}.
Yet, we slightly modify their approach to obtain a simpler reflection law. 

As in \cite{Rabitti2014}, we consider a wave beam of frequency $\omega$ that is sufficiently localised such that it travels along  a geometrical ray path. 
Each ray is characterised by its angle of propagation $\varphi$ with respect to the vertical rotation axis $\bm{e}_z$ and by its azimuthal angle $\phi$ with respect to the axis $\bm{e}_x$. 
We assume $0\leq \varphi \leq \pi$ and $0\leq \phi < 2 \pi$. 
The angle $\varphi$ is given by $\pi/2 \pm \theta$ where the angle $\theta$ (between 0 and $\pi/2$) is fixed by the frequency of the inertial oscillations according to the dispersion relation of inertial waves, which in our dimensionless formulation is simply
\begin{equation}
    \omega=\cos\theta \ .
\end{equation}
This condition means that the rays propagate in an axial cone that makes an angle $\theta$ with respect to the horizontal plane.   
This property is maintained  during the ray propagation including when it reflects on solid boundaries \citep{Phillips_book}.
In two-dimensional or axisymmetric domains, this non-specular reflection can lead to limit cycles towards which all rays converge.
These cycles are called attractors.

The horizontal (or azimuthal) angle $\phi$ is irrelevant in two dimensions since the ray always propagates inside the same predefined plane.
In three dimensions however, the ray is free to move through the whole volume.
Contrary to $\varphi$, $\phi$ is not conserved during reflections on solid boundaries.
The reflection law is not specular but corresponds to a tendency for the ray to converge towards the vertical plane containing the steepest descent direction.
This has been extensively discussed in \cite{Manders2004}, \cite{Rabitti2013,Rabitti2014} and \cite{Pillet2018,Pillet2019}. 

\begin{figure}
\centering
(a)\includegraphics[width=0.45\textwidth]{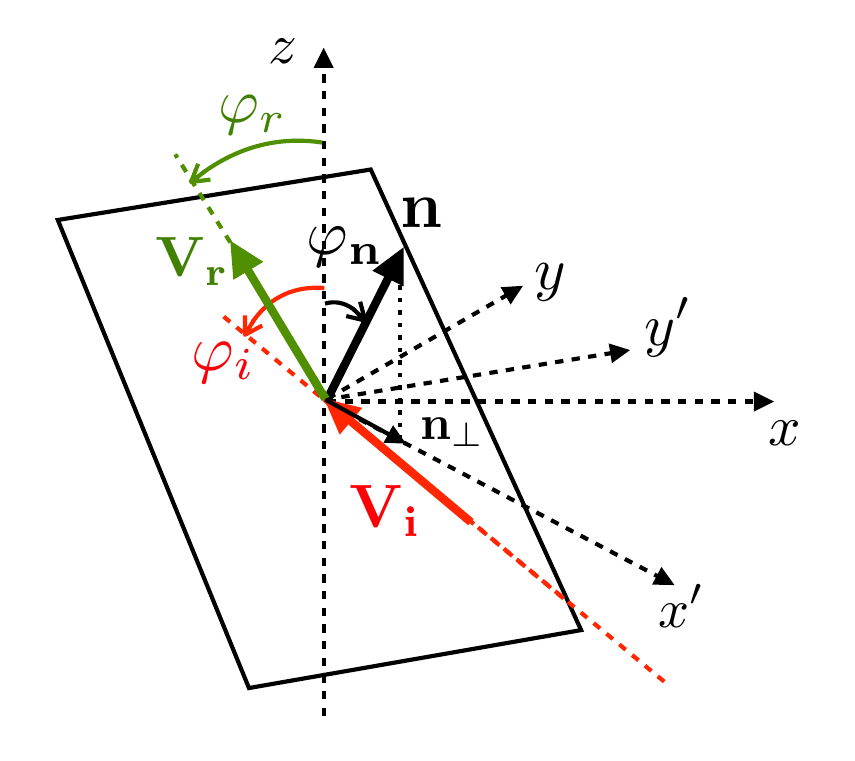}\hfill
(b)\includegraphics[width=0.45\textwidth]{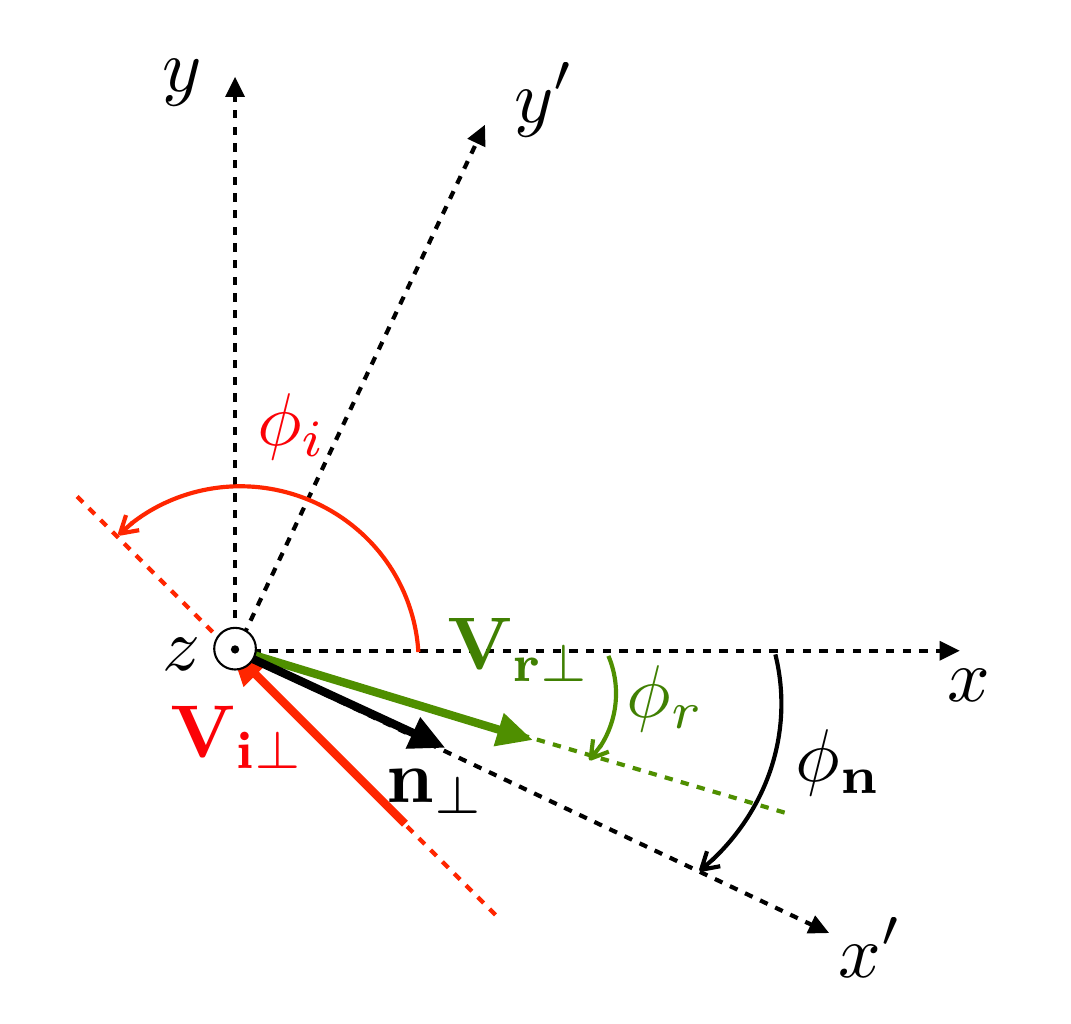}
 \caption{Reflection of an inertial beam on an inclined surface. The incident and reflected beams are defined by their velocity vectors ${\bf V_i}$ and ${\bf V_r}$, the surface by its normal vector ${\bf n}$ oriented toward the fluid. (a) 3D view; (b)  Projected view on the horizontal plane.}
 \label{fig:reflection}
\end{figure}

In the present study, we shall  use the following relation  between the incident and reflected angles $\phi_i$ and $\phi_r$ when a ray reflects on a plane surface 
inclined by an angle $\alpha$ (between 0 and $\pi/2$) with respect to the horizontal plane (see figure \ref{fig:reflection})
\be
\tan (\phi_r -\phi_{\bf n}) = \frac{(\tan^2\theta - \tan^2\alpha) \sin(\phi_i -\phi_{\bf n})}{  (\tan ^2\theta + \tan^2 \alpha ) \cos(\phi_i -\phi_{\bf n}) + 2 \zeta  \tan \alpha \tan \theta } \ .
\label{exp:tanphir}
\ee 
In this expression, $\phi_{\bf n}$ is the azimuthal angle of the normal vector ${\bf n}$ of the surface oriented towards the fluid and  $\zeta = {\rm sgn}(n_zV_{iz})$, where $n_z$ is the vertical component of the normal vector and $V_{iz}$ is the vertical component of the incident velocity.
A similar expression was derived in  \cite{Manders2004} for wavevector angles in the context of plane wave reflection.
In term of wave propagation directions, the reflection law has never been written 
in this form in the literature.  We provide a derivation  in Appendix A.

\subsection{Geometry of the fluid domain}

We consider the volume contained within a truncated elliptic cone defined by
\be 
x^2 + \frac{y^2}{b^2} \leq \frac{z^2}{\tan^2 \alpha}  ~,
\label{eq:cone}
\ee
with $\tan\alpha\leq z \leq \tan\alpha+H$.
The angle $\alpha$ is the angle made by the conical surface with respect to the horizontal in the $x$-direction.
The base of the cone always has a unit radius along the $x$-axis while it reaches $b$ along the $y$-axis.
Without loss of generality, we assume $b\geq1$ so that the short axis is always along the $x$-axis.
As it will become apparent later, the attractor will localise in that case onto the particular plane $(Oxz)$.
The case $b=1$ corresponds to the classical axisymmetric truncated cone geometry.
This type of axisymmetric geometries involving a truncated conical surface, sometimes called a frustum, has already been studied previously \citep{beardsley1970experimental,henderson1992finite,borcia2013inertial,Klein2014,sibgatullin2019influence,pacary2023}.
The novelty of our study is to extend the linear wave dynamics to the case of non-axisymmetric domains which correspond in our case to $b>1$.
Although theoretical results will be derived for any values of $H$ and $\alpha$, the numerical investigations will focus on the particular parameters $H=1$ and $\alpha=\pi/4$ and consider super-critical slopes only for which $\theta<\alpha$.

\section{Ray tracing and local analysis}
\label{sec:tracing}

In this section, we discuss the properties of the three-dimensional ray paths satisfying the reflection laws discussed above.
We start by discussing the axisymmetric case ($b=1$) before considering ray paths in the non-axisymmetric geometry ($b>1$).

\subsection{Axisymmetric truncated cone} 

In this section, we consider the case $b=1$ in equation~\eqref{eq:cone} which corresponds to an axisymmetric truncated cone. 
This geometry has been recently considered in \cite{pacary2023}.
It corresponds to the axisymmetric version of the 2D trapezoidal geometry that has been studied in
numerous works, as described in the introduction. 
Owing to the axisymmetry, the normal vector of the conical surface is always oriented toward the axis of symmetry $Oz$.
This implies that if a ray is oriented towards this axis, it reaches the conical surface with a meridional angle $\phi_i = \phi_{\bf n}+ \pi$, and reflects with an angle $\phi_r =\phi_{\bf n}$ as prescribed by \eqref{exp:tanphir}.
It therefore continues to be oriented towards the axis.
This means that if a ray initially lies within a meridional plane, it remains confined to this plane forever as in a 2D geometry. 

\begin{figure}
    \centering
    (a)\includegraphics[width=0.48\textwidth]{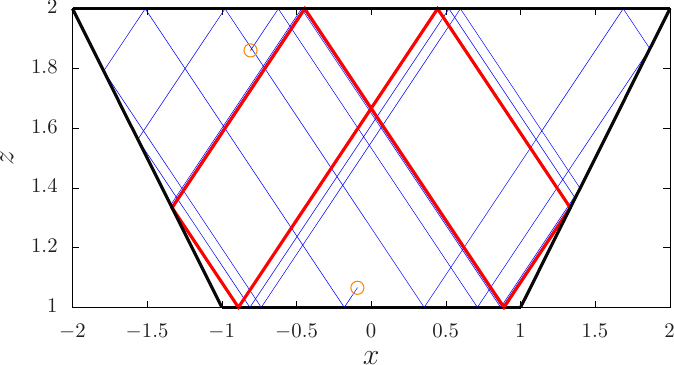}(b)\includegraphics[width=0.44\textwidth]{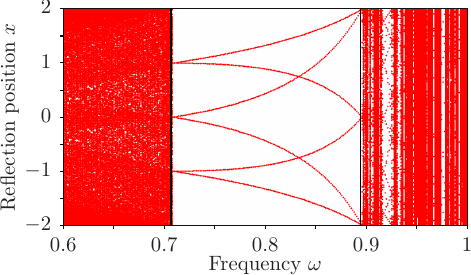}
    \caption{(a) 2D ray paths for two particular trajectories starting from the orange empty circles and for $\omega=0.8$. The red path corresponds to the last half of the many reflections and shows the final attractor. (b) Position of each reflection along the $x$ axis of the last few reflections as a function of $\omega$. The two vertical lines correspond to $\omega=1/\sqrt{2}$ and $\omega=2/\sqrt{5}$.}
    \label{fig:2d}
\end{figure}

The ray paths confined within a particular meridional 2D plane of the axisymmetric cone, which is equivalent to a 2D trapezoid, have been analysed in many papers \citep{Maas1995,Maas1997,Grisouard2008,brunet2019linear,pacary2023}.
They converge towards attractors owing to the focusing effect resulting from the reflection along the conical slope.
This is illustrated in figure~\ref{fig:2d}(a) where we show two ray paths converging towards an attractor for the particular case $\omega=0.8$, $H=1$ and $\tan\alpha=1$. 
This particular attractor (shown in red in figure~\ref{fig:2d}(a)) is composed of two symmetrical quadrangles and is the simplest attractor (with the lowest number of reflections) that can be obtained in this geometry. 
It exists for $\alpha > \theta$ (\textit{i.e.} when the reflection on the sloping boundary is subcritical) and in a finite range of parameters that can be obtained by finding the coordinates of the reflection points. 
For instance, the points $(\pm x_a,z_a)$ where the attractor reflects on the inclined boundary (see figure~\ref{fig:Geometry-3D} below) can be expressed in terms of 
the angles $\alpha$ and $\theta$ and the height $H$ as
\be
x_a = \frac{H}{\tan\theta} ~,~~ z_a = \frac{H \tan \alpha}{\tan \theta} \ .
\label{exp:ra}
\ee
Writing the condition that $z_a$ is between $\tan \alpha$  and $\tan \alpha + H$, 
implies the following condition
\be 
\tan \theta < H < \frac{\tan\alpha \tan\theta}{\tan \alpha - \tan \theta} \ .
\ee
This condition can be written in term of the frequency $\omega =\cos \theta$ as
\be
\frac{1}{\sqrt{1 + H^2} }< \omega <\frac{\tan\alpha + H}{\sqrt{ (\tan\alpha + H)^2 + H^2\tan^2\alpha}} 
\label{exp:omega-attractor}
\ee
which defines the frequency range for which this type of attractor exists. 
For the parameters used in figure \ref{fig:2d} ($H=1$ and $\tan\alpha =1$), it corresponds to the interval $1/\sqrt{2}< \omega < 2 \sqrt{5}$. 
This interval can be seen in figure~\ref{fig:2d}(b) where we show the positions on the $x$-axis of each reflection (after a large number of reflections in order to focus on the attractor path) for many random initial conditions uniformly distributed across the surface and as a function of frequency.
There are no attractors for frequencies below $\omega<1/\sqrt{2}$.
The relatively empty regions for these low frequencies correspond to rays trapped in the upper corners.
For $\omega>2/\sqrt{5}$, other attractors are still observed but they are now characterised by a more complex path involving multiple reflections on each of the boundaries.
In the following, we will focus on the simpler attractor observed for frequencies satisfying the conditions \eqref{exp:omega-attractor}.

\begin{figure}
    \centering
    (a)\includegraphics[width=0.58\textwidth]{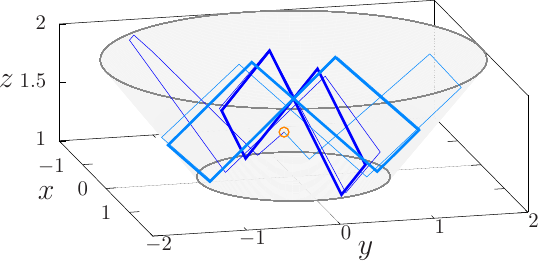}\hfill(b)\includegraphics[width=0.3\textwidth]{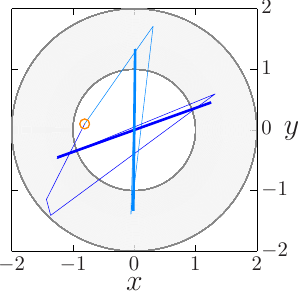}
    \caption{Ray path for two particular trajectories initialised at the same point indicated by the empty circle. The two trajectories differ only by the initial horizontal angle of propagation. The thin lines correspond to the transient propagation while the thick lines correspond to the final limit cycle after many reflections. The conical surface is coloured in light grey to distinguish it from the horizontal planes. The height of the cone is $H=1$, its opening angle is $\tan\alpha=1$ and the frequency is $\omega=0.8$. (a) Side view. (b) View from the top. A movie (Movie1) showing the propagation of many randomly initialised rays can be found in Supplementary materials.}
    \label{fig:axi}
\end{figure}

If we now authorise the ray beams to deviate from a particular meridional plane, their path becomes more complex.
One has to monitor the horizontal angle $\phi$ of the direction of the ray and the meridional angle $\psi$ of the position  where it reflects on the boundaries (see figure~\ref{fig:Geometry-3D}(b) below). 
This problem has been recently studied in \cite{pacary2023} for this particular geometry (see also section 7 in \cite{maas2005wave}).
They showed that the 2D attractor is still obtained but its location along the azimuthal direction is now dependent on the choice of initial conditions.
This phenomenon is illustrated in figure~\ref{fig:axi} where we show that two rays emitted from the same location with different initial horizontal angles of propagation $\phi$ end on the same 2D attractor but in different meridional planes.
However, the geometry being azimuthally invariant, all the meridional planes are possible and should be obtained  with the same probability.
A fully axisymmetric attractor would then be obtained (with a non-uniform distribution of trapping planes though, see \cite{maas2005wave}) if all the possible initial conditions were simultaneously considered (with the exception of whispering gallery modes \citep{Pillet2019} which we did not observe here).
More details can be found in \cite{Pacary_thesis,pacary2023}, where such an axisymmetric system is explored using both ray tracing and experiments.

\subsection{Elliptic cone}

In this section, we consider the unexplored case $b>1$ in equation~\eqref{eq:cone} which corresponds to a truncated elliptic cone.
As the axisymmetry is now broken, we do not expect the presence of any axisymmetric attractor. 
In the following, we focus on the case where the reflection on the conical surface remains subcritical which leads to the following upper bound on $b$
\begin{equation}
\label{eq:bc1}
b<b_c=\frac{\tan\alpha}{\tan\theta} \ .    
\end{equation}

The first effect of the elliptic deformation is to modify the orientation of the normal vector of the conical surface.
Except in the vertical planes $x=0$ and $y=0$ corresponding to the directions of the principal axes of the elliptical cone, the normal vector to the cone is no longer oriented towards the vertical rotation axis.
This means that contrary to the axisymmetric case, no ray can stay trapped in a particular meridional plane apart from the planes $x=0$ and $y=0$.
When a ray is initialised inside any other meridional plane, it deviates from it after its first reflection on the conical surface. 
Whether the planes $x=0$ and $y=0$ are stable or unstable equilibrium will be discussed below.

\begin{figure}
    \centering
    (a)\includegraphics[width=0.46\textwidth]{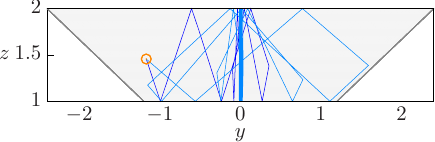}\hspace{1mm}(b)\includegraphics[width=0.46\textwidth]{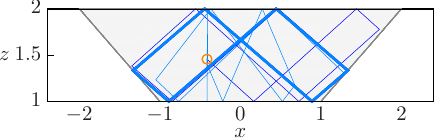}\\
    (c)\includegraphics[width=0.62\textwidth]{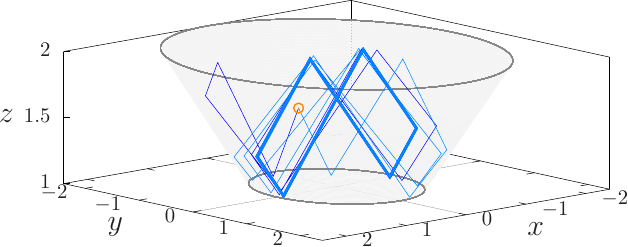}\hspace{1mm}(d)\includegraphics[width=0.3\textwidth]{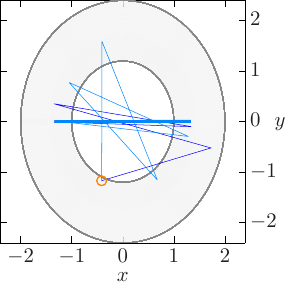}
\caption{Same as figure~\ref{fig:axi} but for a non-axisymmetric domain with $b=1.2$ in equation~\eqref{eq:cone}. The other parameters are $\omega=0.8$, $H=1$ and $\tan\alpha=1$. (a) Side view in the $(y,z)$ plane. (b) Side view in the $(x,z)$ plane. (c) 3D view. (d) Top view in the $(x,y)$ plane. A movie (Movie2) showing the propagation of many randomly initialised rays can be found in Supplementary materials.}
    \label{fig:nonaxi}
\end{figure}

\begin{figure}
    {\centering
    (a)\includegraphics[width=0.46\textwidth]{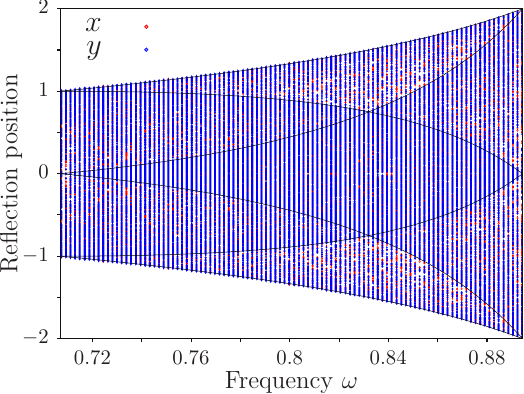}(b)\includegraphics[width=0.46\textwidth]{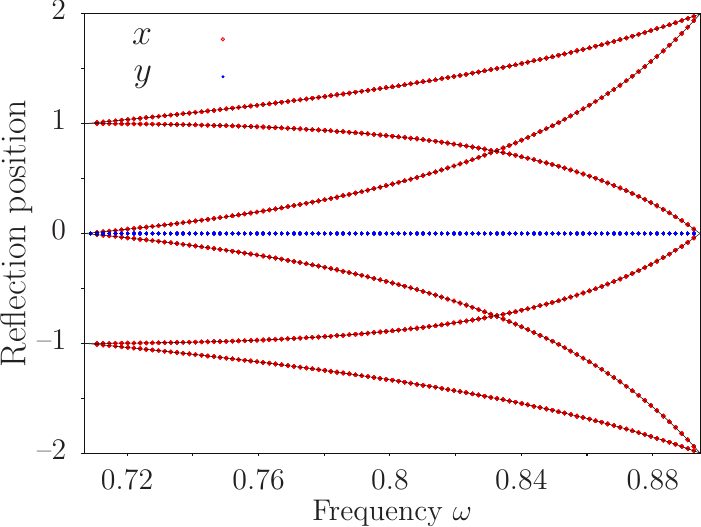}}
\caption{Positions of the reflection points along the $x$ axis in red and along the $y$ axis in blue for the final limit cycle obtained from many random initial conditions. (a) Axisymmetric case $b=1$ and (b) elliptic case $b=1.2$. The cone height is $H=1$ and its opening angle is $\tan\alpha=1$. The graph focuses on the frequency range defined by equation~\eqref{exp:omega-attractor}. The thin lines correspond to the analytical position of the six reflection points of the 2D attractor.\label{fig:po3d}}
\end{figure}

We first repeat the ray tracing experiment done previously for $b=1$ but in a non-axisymmetric geometry with $b=1.2$.
The same frequency $\omega=0.8$ is considered.
An example is shown in figures~\ref{fig:nonaxi}, where two ray paths are generated from the same initial position but with different horizontal angles.
We observe that the two ray paths now converge towards the same unique attractor localised in the plane $y=0$ containing the semi-minor axis of the elliptical cone.
One can repeat the experiment for many initial positions and horizontal angles $\phi$ showing that all rays eventually end up in this particular plane $y=0$ and along the same limit cycle.
For this particular case, it thus seems that the plane $y=0$ is a stable equilibrium while the plane $x=0$ is unstable.
This is further confirmed in figure~\ref{fig:po3d} where we show the positions $x$ and $y$ of the reflection points of the limit cycle (obtained by considering the last ten iterations of a total of five hundred) for many frequencies and many random initial conditions.
We compare the axisymmetric case $b=1$ in figure~\ref{fig:po3d}(a) with the elliptic case $b=1.2$ in figure~\ref{fig:po3d}(b).
The thin lines indicate the $x$ coordinates of the six reflection points of the 2D attractor path, as already shown in Figure~\ref{fig:2d}(b), in the particular plane $y=0$ which is the same for all ellipticities $b$ in our case.
For the axisymmetric case, each independent ray path converges towards a similar 2D attractor in a particular meridional plane, depending on its initial conditions, leading to a dense pattern of reflection points when plotting their $(x,y)$ coordinates.
Note that the projection from the axisymmetric domain to the arbitrary coordinates $x$ and $y$ leads to a non-uniform density of reflection points which tend to accumulate close to the expected positions of the 2D attractors (shown by the thin lines).
For the non-axisymmetric case $b=1.2$ however, all the rays converge towards the particular plane $y=0$, regardless of their initial conditions in terms of position and initial horizontal angle, while we recover the same attractor structure as in 2D when considering the $x$ coordinates.
Note that we recover this peculiar property for all frequencies within the attractor range given by equation~\eqref{exp:omega-attractor}.

The attractor in the particular plane $y=0$ seems to attract all the rays, regardless of their initial positions within the volume, at the exception of rays initialised within the $x=0$ plane which is also an equilibrium, albeit unstable.
Contrary to the axisymmetric case, there is a second azimuthal convergence (related to the varying curvature of the elliptical cone along the azimuthal direction) in addition to the meridional convergence at the origin of the classical 2D attractor (related to the inclination of the conical surface).
For this reason, we call this final limit cycle a ``super-attractor", to differentiate it from the classical attractor surface observed in the axisymmetric case.
The super-attractor results from a convergence of rays in both axial and azimuthal directions while there is no azimuthal convergence for an attractor in an axisymmetric geometry, as its azimuthal position depends on the initial position and horizontal angle of propagation.

\subsubsection{Local analysis of the super-attractor\label{sec:local}}

In this section, we analyse the local properties of the super-attractor.
Our objective is to confirm the attracting character of the super-attractor both towards a particular limit cycle in the meridional plane and towards a particular meridional plane regardless of initial conditions.
While the first property is shared with regular attractors, we show that the second is specific to super-attractors.
We aim to obtain an analytic expression for the attraction rate (i.e. Lyapunov coefficient) as a function of the geometrical parameters. 

We focus on the super-attractor which is made of two symmetrical quadrangles as illustrated in 
figure \ref{fig:2d}(a).  This 2D attractor has been studied in section \S 3.1. It exists in the plane $x=0$ for 
frequencies in the interval defined in (\ref{exp:omega-attractor}). 
The points $(\pm x_a,0,z_a)$ where reflection occurs has been given in (\ref{exp:ra}).

\begin{figure}
    \centering
    (a)\includegraphics[width=0.42\textwidth]{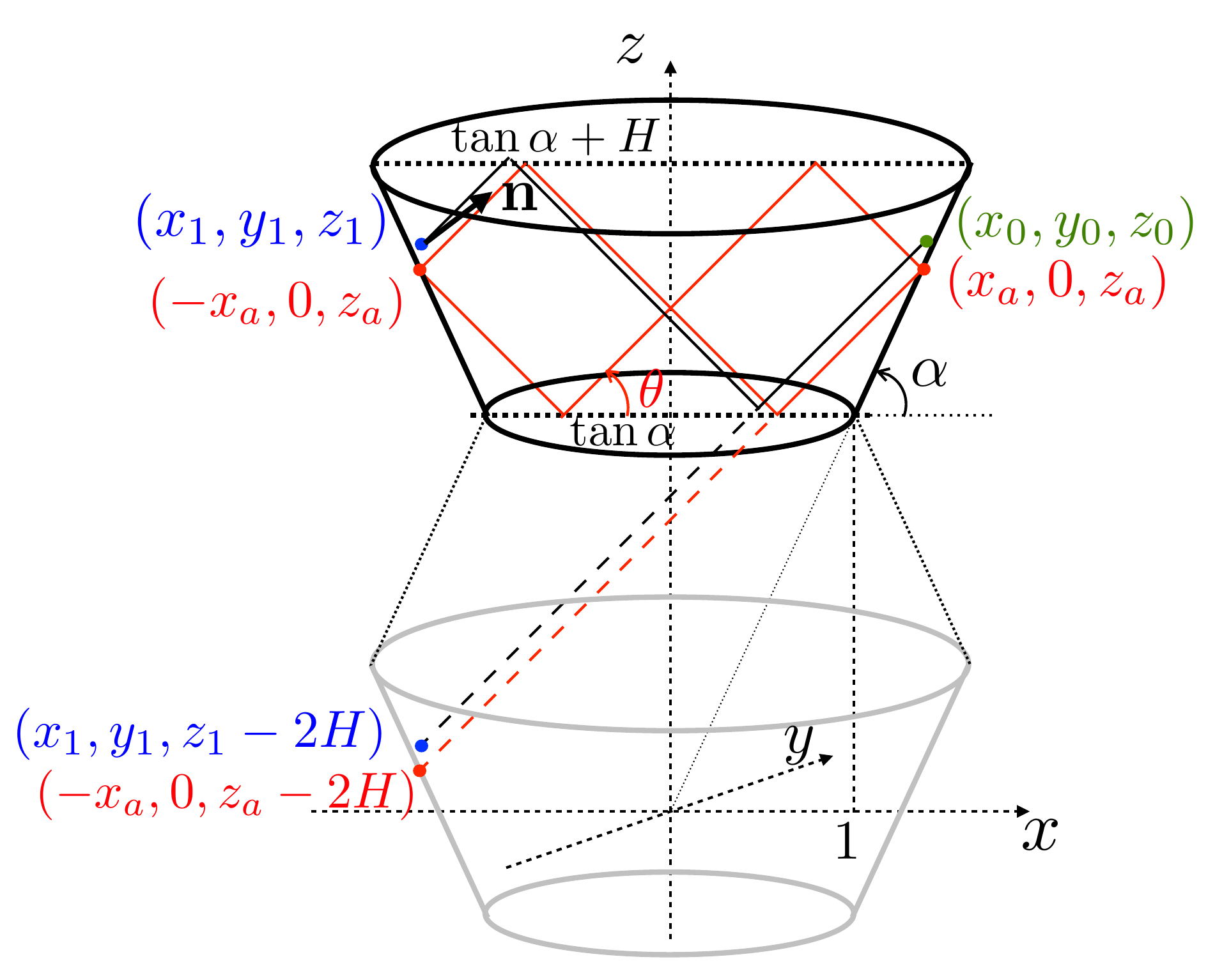}(b)\includegraphics[width=0.5\textwidth]{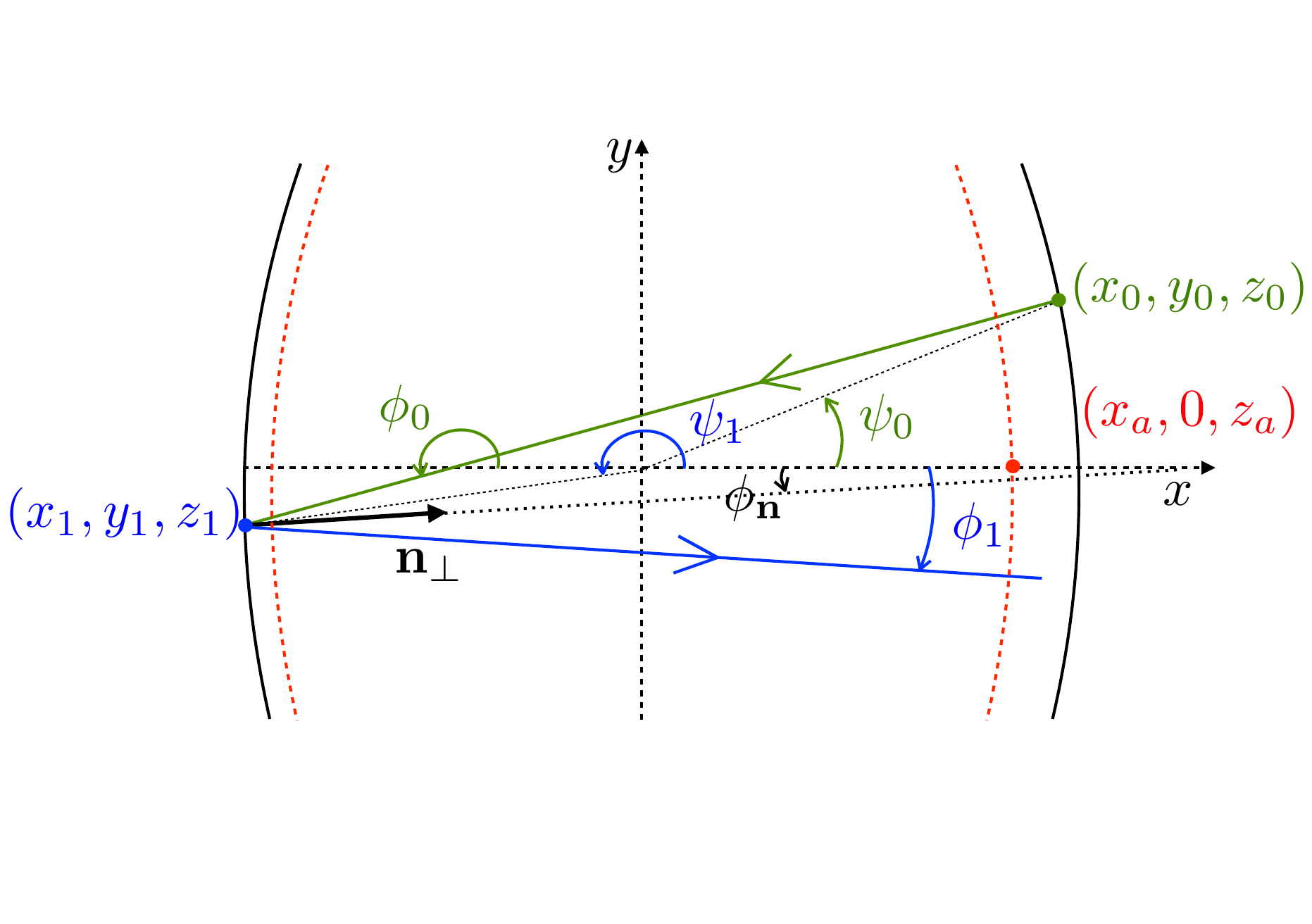}
    \caption{(a) Perspective view of the elliptical cone in thick solid line. The attractor is indicated in red line. 
    The cone is twice symmetrically duplicated to avoid considering the reflections on the horizontal planes. (b) View from the top of a ray emitted close to the attractor. The red dashed lines indicate the section of the cone at the altitude $z_a$ of the attractor. The ray emitted from the point $(x_0,y_0,z_0)$
   with a horizontal angle $\phi_0$ is reflected at $(x_1,y_1,z_1)$ with the angle $\phi_1$.}
    \label{fig:Geometry-3D}
\end{figure}

We now want to analyse the behaviour of a ray close to the attractor. 
We consider a ray emitted from a point $(x_0,y_0,z_0)$ on the inclined surface close to $(x_a,0,z_a)$. This ray is emitted downward with a vertical angle $\varphi_0 = \pi/2 + \theta$ and an azimuthal angle
$\phi_0$ as illustrated in figure \ref{fig:Geometry-3D}.
If this ray is not too far from the attractor, it first reflects on the lower plane surface, then reflects on the upper plane surface before reaching 
the inclined surface on the other side in a point $(x_1,y_1,z_1)$ close to $(-x_a,0,z_a)$. 
Both points $(x_0,y_0,z_0)$ and $(x_1,y_1,z_1)$ are on the cone defined by \eqref{eq:cone} so their horizontal coordinates can be written as 
\bsea
 && x_0 =  z_0 \frac{\cos \psi_0 }{\tan \alpha}  ~,~~ y_0 = b z_0 \frac{\sin \psi_0 }{\tan \alpha} , \\
 &&  x_1 =  z_1 \frac{\cos \psi_1 }{\tan \alpha}  ~,~~ y_1 = b z_1 \frac{\sin \psi_1 }{\tan \alpha} , 
 \label{exp:x0}
 \esea
where  $\psi_0$ and $\psi_1 $ are  their azimuthal angles (see figure \ref{fig:Geometry-3D}(b)).

As the reflections on the plane surfaces do not modify the azimuthal angle of the ray (see equation~\eqref{exp:tanphir} for the particular case $\alpha=0$), 
the coordinates of the point $(x_1,y_1,z_1)$ can be obtained by continuing the ray in horizontal mirror images of the cones. Such a ray reaches the boundary of the second cone image in a point which has just been shifted vertically by twice  the height of the cone, that is in $(x_1,y_1,z_1-2H)$ (see figure \ref{fig:Geometry-3D}). 
This property means that there exists $\lambda$ such that
\bsea 
&&x_1 = x_0  - \lambda \cos \theta \cos \phi_0 \ , \\
&&y_1 = y_0 - \lambda \cos \theta \sin \phi_0 \ ,\\
&&z_1 -2H = z_0 + \lambda \sin\theta \ .
\label{exp:x1}
\esea

When the ray is close to the attractor, $(x_0,y_0,z_0)$ and $(x_1,y_1,z_1)$ are assumed to be close to $(x_a,0,z_a)$ and $(-x_a,0,z_a)$, respectively. 
Moreover, $\psi_0$ is assumed to be small, and $\phi_0$ and $\psi_0$  are close to $\pi$. 
At first order, equations (\ref{exp:x0})  give for the relative distances
\bsea  
&& \delta x_0 \sim   \frac{\delta z_0 }{\tan \alpha}  ~,~~ \delta y_0 \sim b z_a \frac{ \psi_0 }{\tan \alpha} \ , \\
 && \delta  x_1 \sim  - \frac{\delta z_1 }{\tan \alpha}  ~,~~ \delta y_1 \sim - b z_a \frac{ (\psi_1 - \pi) }{\tan \alpha} \ .
\label{exp:deltax0}
\esea
From (\ref{exp:x1}c), one gets $\lambda$ as
\be
\lambda \sim \frac{-2H + \delta z_1 -\delta z_0}{\sin\theta} \ .
\ee
Inserting this expression in (\ref{exp:x1}a), one obtains using (\ref{exp:deltax0}a,b)
\be 
\delta x_1 \sim  - K \delta x_0  ~,~~ \delta z_1 \sim K \delta z_0 \ ,
\ee
 with 
\be 
K = \frac{\tan \alpha - \tan \theta}{\tan\alpha + \tan \theta} = \frac{\sin(\alpha- \theta)}{\sin(\alpha+\theta)} \ ,
\label{exp:K}
\ee
which corresponds to the 2D contraction factor of the attractor. 

In (\ref{exp:x0}b), we obtain 
\be 
\delta y_1 \sim \delta y_0 - \frac{2 H (\phi_0 -\pi)}{\tan\theta} ,
\label{exp:deltay1}
\ee
 which gives a first relation between the angles
 \be 
 \psi _1 - \pi = -\psi _0 + \frac{2 }{b} (\phi_0 - \pi) ~.
 \label{exp:psi1}
 \ee
 
A second relation that expresses the angle $\phi_1$ of the reflected ray in terms of $\phi_0$ and  $\psi_0$
is obtained by applying the condition of reflection (\ref{exp:tanphir}) at $(x_1,y_1,z_1)$.
Close to the attractor, the azimuthal angle $\phi_{\bf n}$ of the normal vector ${\bf n}$ is given 
at leading order by
\be
\phi_{\bf n} \sim \frac{(\psi_1 - \pi)}{b}  ~.
\label{exp:phin}
\ee
The normal vector ${\bf n}$ and the incident rays are oppositely oriented with respect to the vertical, so $\zeta ={\rm sgn}(n_z V_{iz}) = -1$. 
Expression (\ref{exp:tanphir}) then gives for small angles
\be
\phi_1  - \phi_{\bf n}  \sim  -\frac{\tan \alpha - \tan \theta}{\tan\alpha + \tan \theta} (\phi_0 - \pi - \phi_{\bf n}) = - K (\phi _0 -\pi -\phi_{\bf n}) .
\ee 
 Using (\ref{exp:psi1}) and (\ref{exp:phin}), this expression finally reduces to
 \be
 \phi_1 \sim -\frac{1+K}{b} \psi_0 + \left(\frac{2(1+K)}{b^2} - K\right) (\phi_0 - \pi) ~.
\label{exp:phi1}
 \ee
 
Expressions (\ref{exp:psi1}) and (\ref{exp:phi1}) can be written as
\be
\bPsi_1 =  \bcM \bPsi_0 
\ee   
 for the vectors $\bPsi_1 = (\psi_1 - \pi, \phi_1)^\top$ and $\bPsi_0 = (\psi_0, \phi_0 -\pi)^\top$ 
where the matrix $ \bcM$ is
\be
 \bcM = \left( \ba{cc} -1 & 2/b \\[0.1cm] -(1+K)/b  & 2(1+K)/b^2 - K \ea \right) .
\ee
The operation can be repeated and after $2N$ reflections on the cone boundary, that is $N$ cycles, we get 
\be 
\bPsi_{2N} = \bcM^{2N} \bPsi_0 
\ee
where $\bPsi_{2N} = (\psi_{2N}, \phi_{2N} - \pi)^\top$.
 
Introducing the eigenvalues $\lambda _{\pm}$ and associated  eigenvectors $\bPsi_{\pm}$ of the matrix  $\bcM$
which are defined by
 \bsea
&&  \lambda _{\pm} = \frac{(1+K)(2/b^2-1) \pm \left((1+K)^2 (2/b^2-1)^2 - 4K\right)^{1/2}}{2} \ ,
\label{exp:lambdapm}
\\
 &&\bPsi_\pm = (2/b, 1 + \lambda_{\pm})^\top \ ,
\label{exp:psilambdapm}
\esea 
$\bPsi_{2N} $ can be written  as
\be
\bPsi_{2N} = C_0 (\lambda _+)^{2N}  \bPsi_{+}  + D_0 (\lambda _-)^{2N}  \bPsi_{-} \ ,
\ee
  where $C_0$ and $D_0$ are constants depending on the initial condition only
  \bsea
&&  C_0= \frac{2(\phi_0-\pi) - b(1+\lambda_-)\psi_0}{2(\lambda_+ -\lambda_-)} \ ,\\
&&  D_0= \frac{2(\phi_0-\pi) - b(1+\lambda_+)\psi_0}{2(\lambda_- -\lambda_+)} \ .
\label{exp:C0D0}
   \esea

The functions $\lambda_+$ and  $\lambda_-$ characterise the behaviour of the angles $\psi_{2N}$ and $\phi_{2N} -\pi$ 
as a function of the cycle number $N$. These angles go to zero if and only if  $|\lambda_\pm| <1$. 
This condition is here equivalent to $b>1$ (since $0<K<1$). 
The functions $\lambda_{\pm}$ depend on the value of $b$ with respect to the particular values
\be
\label{bcp}
b_{c\pm} = \frac{\sqrt{2(1+K)}}{1\pm \sqrt{K}} \ .
\ee

The two $\lambda_\pm$ are real positive for $1<b<b_{c+}$, and real negative for $b>b_{c-}$. For $b_{c+}<b<b_{c-}$, they are complex conjugates with a constant absolute values equal to $\sqrt{K}$.  
When $b=b_{c\pm}$, the solution evolves differently as shown in appendix B.

\subsubsection{Comparison between local analysis and numerical ray tracing}

This section compares the prediction of the local analysis with that of the global ray tracing approach described above.
To do so, we randomise the initial position inside the volume, the horizontal angle of propagation and the sign of the vertical velocity component.
We then track the ray for $10^4$ reflections on boundaries which is enough to reach the attractor in most cases.
The horizontal angle is computed at each reflection on the conical surface and for $x_i>0$ according to $\psi_{i}=\textrm{arg}(x_i+\textrm{i}y_i)=\textrm{atan2}(y_i,x_i)$ where $(x_i,y_i)$ are the coordinates of the $i^{\textrm{th}}$ reflection point on the positive half $x>0$ of the conical surface.
The horizontal angle difference $\Delta\psi_i=|\psi_i-\psi_{i-1}|$ is then tracked as a function of the number of cycles.
Note that we track the convergence of the horizontal angle after a complete cycle around the attractor for comparison with the prediction~\eqref{exp:lambdapm} of the local analysis presented in section~\ref{sec:local}.
In that case, the analysis is performed on the number of cycles around the attractor and not on the number of reflections on the boundaries (there are 6 reflections on boundaries and 2 reflections on the conical surface per cycle for the particular attractor considered here).

\begin{figure}
    \centering
    (a)\includegraphics[width=0.465\textwidth]{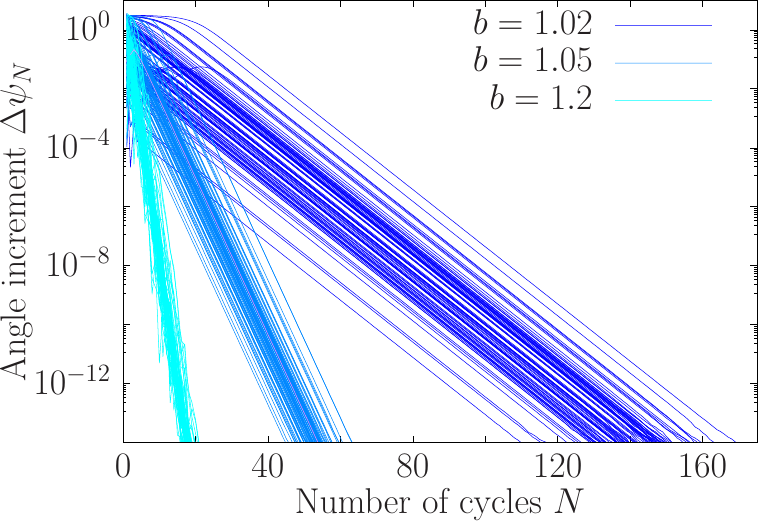}
    (b)\includegraphics[width=0.45\textwidth]{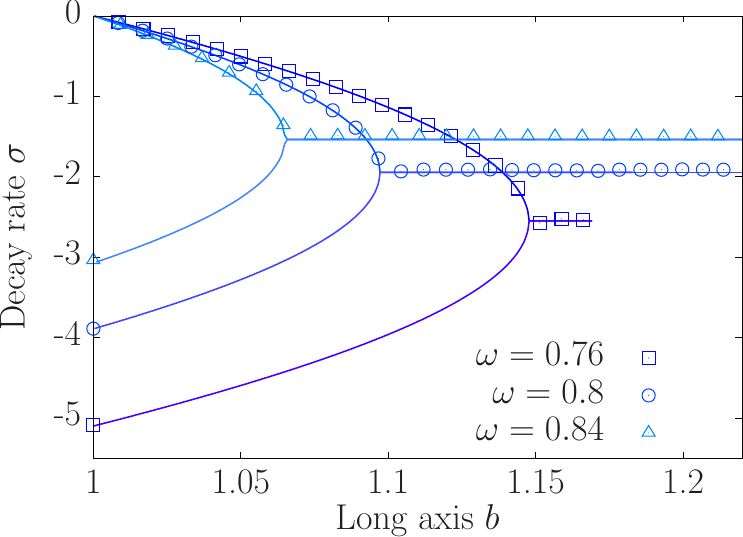}
    \caption{(a) Angle difference $\Delta \psi_N= |\psi_N-\psi_{N-1}|$ in logarithmic scale as a function of the number of cycles $N$. We show 100 independent realisations with different initial conditions for each value of $b$. (b) Decay rate $\sigma$ defined by $\Delta\psi_N\sim \exp(\sigma N)$ as a function of the long axis $b$. Symbols correspond to the ray tracing approach while lines correspond to the linear local analysis \eqref{exp:lambdapm}. The upper branches correspond to $\sigma = \ln|\lambda_+|$ while the lower branches correspond to $\sigma = \ln|\lambda_-|$. Note that the data for $\omega=0.76$ stop because of the upper bound $b_c\approx1.17$ as given by~\eqref{eq:bc1}.}
    \label{fig:conv}
\end{figure}

The evolution of the horizontal angle increment is typically characterised by a short transient followed by an exponential decay until machine accuracy is eventually reached and the ray becomes trapped in a particular plane.
Some examples for $\omega=0.8$, $H=1$, $\tan\alpha=1$ and three values of $b$ are shown in Figure~\ref{fig:conv}(a) to illustrate the convergence process.
We observe that all rays converge towards the super-attractor at the same exponential rate, irrespective of the initial conditions.
The duration of the transient before the ray effectively converges towards the attractor depends on the initial conditions and corresponds to the early reflections far from the final limit cycle.
Note that for this particular case, the particular value of $b$ defined by equation~\eqref{bcp} is $b_{c+}=1.097$.
Consistent with the theoretical prediction, we observe real values for the decay rate when $b<b_{c+}$ (see the two cases $b=1.02$ and $b=1.05$ in figure~\ref{fig:conv}(a)) and complex values when $b>b_{c+}$ (see the case $b=1.2$).

A best fit $\Delta\psi_N\sim \exp(\sigma N)$, where $N$ is the number of cycles, is computed in the range $10^{-14}<\Delta\psi_N<10^{-3}$ and averaged over $10^3$ independent rays initialised randomly within the whole volume.
Results are shown in Figure~\ref{fig:conv}(b).
An excellent agreement between the local theoretical prediction and the ray tracing approach is observed for various values of $b$ and three different frequencies within the attractor range.
Interestingly, the decay rate is $\sigma=\ln|\lambda_-|=\ln|K|$ when $b=1$ while it is smaller when $b>1$.
There is therefore a drastic difference between the case $b=1$, for which all rays converge rapidly toward a plane different for each ray, and the case $b>1$ for which the convergence rate is smaller and actually tends to 0 when $b\rightarrow1$.
Although the breaking of the axisymmetry does create a globally attracting plane as already discussed previously, the rate of convergence towards this unique plane increases with $b$ and is actually maximum once $b>b_{c+}$.
In that case, it is actually half of that observed when $b=1$.

\section{Numerical simulations of the linear viscous problem\label{sec:dns}}

Up to now, we have only discussed the properties of the ray paths which are only valid in the limit of vanishing Ekman numbers.
The link between the properties of the ray paths and the actual viscous solution of the original linear Navier-Stokes equations~\eqref{eq:momentum} is not obvious.
In this context, it is desirable to check whether the globally-attracting solutions discussed in previous sections have any counterpart when considering the direct solution of equations~\eqref{eq:momentum}.

To that end, we solve the linear viscous equations~\eqref{eq:momentum} using the spectral element solver Nek5000 \citep{Fischer1997,Deville2002}.
The domain is discretised using a number $\mathcal{E}$ of hexahedral elements.
Elements have been refined close to all boundaries to properly resolve viscous Ekman layers.
The velocity is discretised within each element using Lagrange polynomial interpolants based on tensor-product arrays of Gauss-Lobatto-Legendre quadrature points.
The polynomial order $l_d$ of the expansion basis on each element is fixed to $11$ in this study, while the number of elements goes up to $\mathcal{E}=29952$ for $E=10^{-7}$.
Convergence has been tested by gradually increasing the polynomial order for a fixed number of elements.
The Coriolis term is treated explicitly by a third-order extrapolation scheme whereas the viscous terms are treated implicitly by a third-order backward differentiation scheme.
Similarly to the ray tracing approach discussed previously, we focus on the particular case $H=1$, $\alpha=\pi/4$ and $\omega=0.8$.

\begin{figure}
    \centering
    (a)\includegraphics[width=0.45\textwidth]{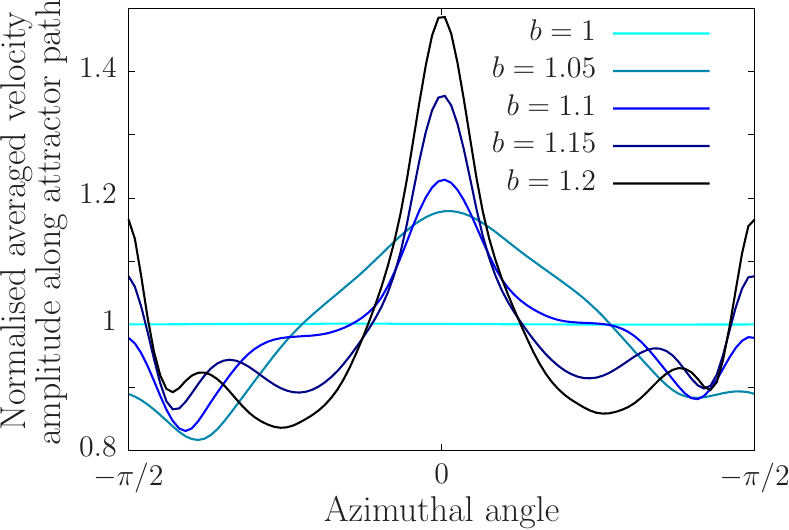}(b)\includegraphics[width=0.46\textwidth]{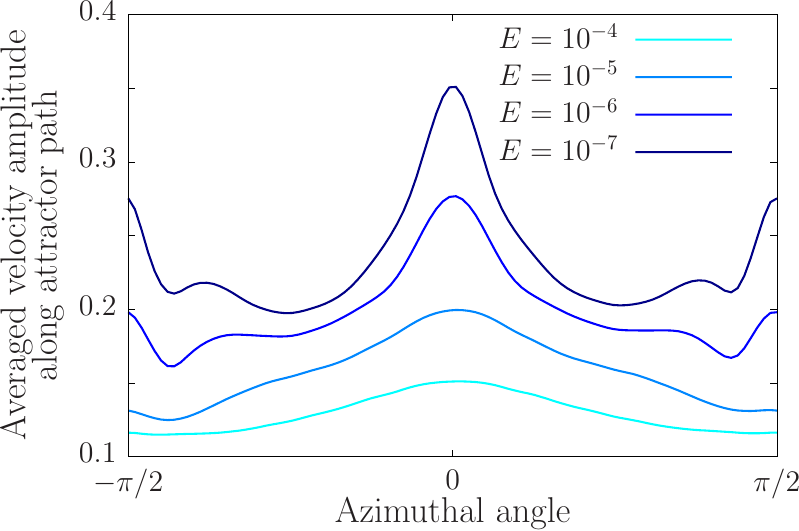}
    \caption{Azimuthal profile of the velocity amplitude averaged over time and over the 2D attractor path on each meridional plane. (a) Variable long axis $b$ at constant Ekman number $E=10^{-7}$. The amplitude is normalised by its azimuthal average. (b) Variable Ekman number at constant long axis $b=1.2$. The forcing frequency is $\omega=0.8$ and the forcing pattern is defined by equation~\eqref{eq:forcing} in all cases.}
    \label{fig:dnsprof}
\end{figure}

\begin{figure}
    \centering
    \includegraphics[width=0.9\textwidth]{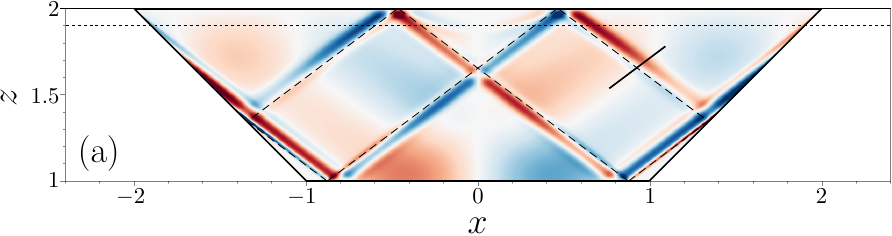}
    \includegraphics[width=0.9\textwidth]{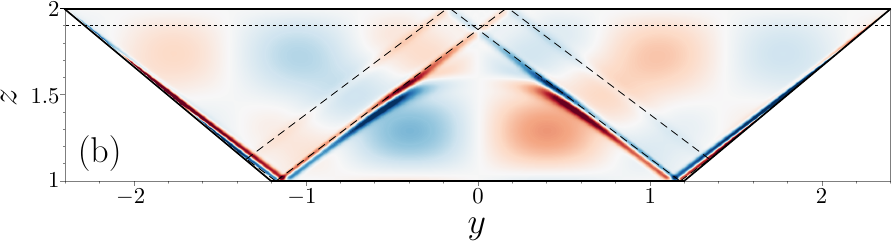}
    \includegraphics[width=0.9\textwidth]{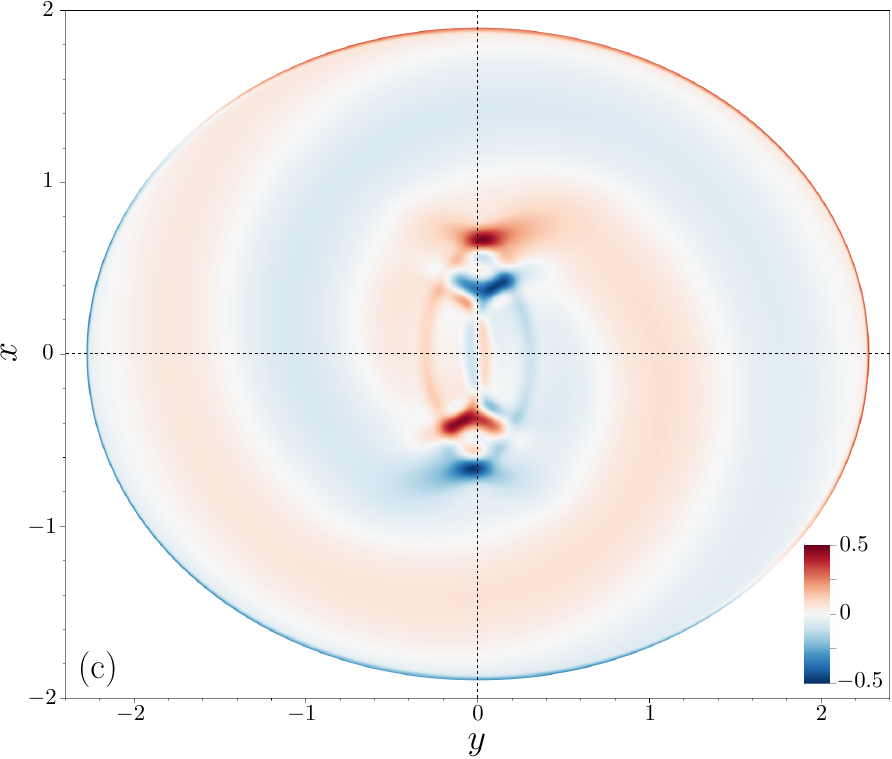}
    \caption{Vertical velocity component in the planes (a) $y=0$, (b) $x=0$ and (c) $z=1.9$. The same color scale is used for the three plots. Parameters are $E=10^{-7}$, $\omega=0.8$ and $b=1.2$. The inclined line in (a) indicates the profile of the plane shown in Figure~\ref{fig:maps}. The dotted lines in (a) and (b) indicate the plane $z=1.9$ shown in (c) while the dash lines show the 2D attractor path in each meridional plane.}
    \label{fig:visit}
\end{figure}

Since we want to compare the axisymmetric conical geometry with its elliptical counterpart, care must be taken when choosing the forcing.
In particular, latitudinal libration is not well suited since the forcing would be of viscous nature in the axisymmetric case (since the velocity at the boundaries would be purely tangential) while there would be a non-zero normal velocity and hence pressure coupling in the non-axisymmetric elliptical case.
Another constraint comes from the corners which will inevitably contribute to the viscous dynamics by emitting their own singular shear layers.
Taking into account these considerations, we opted for a rotating vertical forcing at the bottom plane of the cone, sometimes called negative nutation \citep{sibgatullin2017direct}, defined by
\begin{equation}
\label{eq:forcing}
u_z(z=1)=\begin{cases}
      \big(x\cos(\omega t)-y\sin(\omega t)\big)f(r) & \text{if $r<1$}\\
      0 & \text{if $r>1$}    \end{cases} \ ,
\end{equation}
where $f(r)=2r^3-3r^2+1$ is a smoothing function to ensure that the forcing vanishes close to the corners and $r=\sqrt{x^2+y^2}$ is the cylindrical radius.
The other two velocity components are zero and the other boundaries are all no-slip.
A similar forcing has already been used both experimentally \citep{pacary2023} and numerically \citep{sibgatullin2019influence}.
Note that we obtained qualitatively similar results when considering an axisymmetric forcing similar to that used by \cite{Boury2021} for example.

From a fluid at rest, we run the simulation until a periodic response is obtained.
In order to quantify the inhomogeneities between different attractor planes, we first define the 2D attractor path for each plane obtained from the intersection between a vertical plane containing the origin and the frustum.
Each plane is parametrised with its azimuthal angle with respect to the $x$ direction.
The plane crossing the short elliptical axis thus corresponds to $\phi=0$ while the plane crossing the long elliptical axis corresponds to $\phi=\pm\pi/2$.
Note that the 2D attractor path on each plane depends on the ellipticity $b$.
While the same attractor path is expected on each individual plane when $b=1$, different attractors (but the same rectangular topology) are expected when $b>1$.
We then compute the averaged velocity amplitude along each path by averaging over time once the periodic response is obtained and along the attractor path.
This averaged amplitude is a constant in the axisymmetric case $b=1$ but depends on the orientation of the plane once $b\neq1$.
As we vary the ellipticity of the domain, the amplitude of the response also varies.
In order to focus on the azimuthal inhomogeneities induced by the wave attractor, we further normalise the amplitude by its average over all azimuthal angles $\phi$.

This ratio is plotted in figure~\ref{fig:dnsprof}(a) for a fixed $E=10^{-7}$ and varying $b\geq1$.
As expected, it is unity for the axisymmetric case $b=1$.
As $b$ increases, a clear focusing of the energy along the short axis corresponding to $\phi=0$ is observed.
Note that a second local maximum is also observed along the long axis $\phi=\pm\pi/2$.
The same simulations are now run at a fixed $b=1.2$ and varying Ekman number from $E=10^{-4}$ down to $E=10^{-7}$.
Results are displayed in figure~\ref{fig:dnsprof}(b).
While focusing is observed for all Ekman numbers, it is more and more pronounced as the Ekman number decreases.
Note again that a residual localisation also persists around the long axis $\phi=\pm\pi/2$.
This secondary localisation of the energy is not expected from the ray tracing approach only since it corresponds to an unstable equilibrium as discussed previously in section~\ref{sec:local}.
We have observed such residual localisation strongly depends on the particular choice of forcing and seems to be partially dependent on contributions from the bottom corner (see Figure~\ref{fig:visit}(b)), which goes well beyond our current understanding mostly based on local ray tracing.

\begin{figure}
    \centering
    (a)\includegraphics[width=0.28\textwidth]{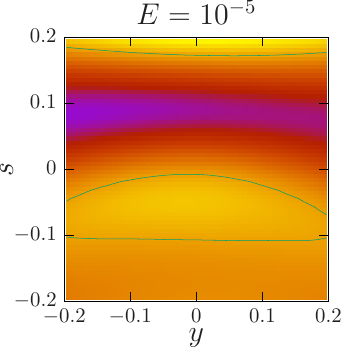}
    (b)\includegraphics[width=0.28\textwidth]{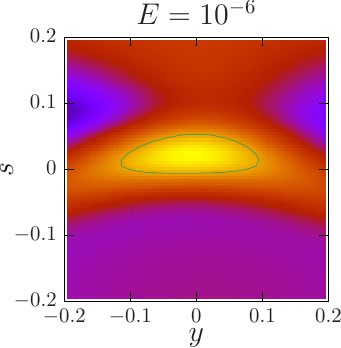}
    (c)\includegraphics[width=0.315\textwidth]{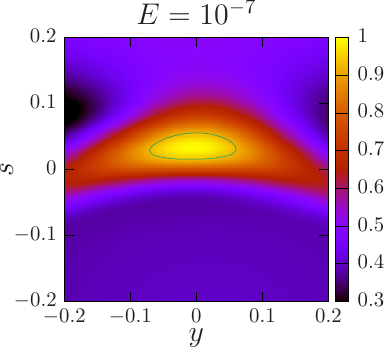}\\
    (d)\includegraphics[width=0.28\textwidth]{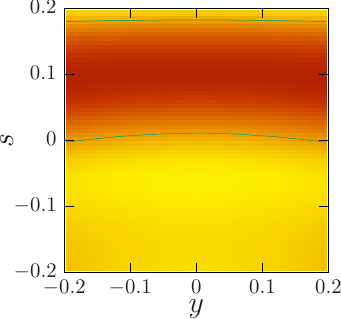}
    (e)\includegraphics[width=0.28\textwidth]{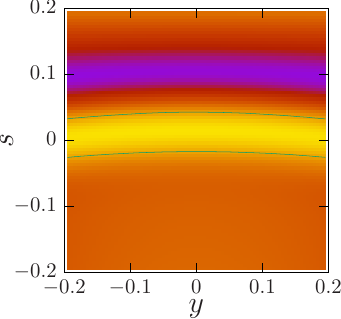}
    (f)\includegraphics[width=0.315\textwidth]{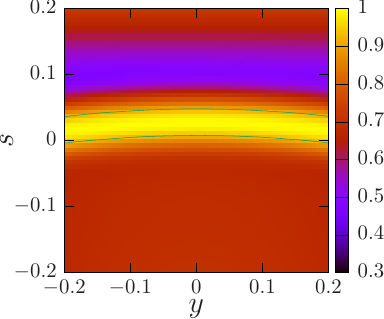}
    \caption{Maps of averaged velocity amplitude normalised by its maximum value on the local plane shown in Figure~\ref{fig:visit}(a). The Ekman number is decreasing from left to right. The top row (a-c) corresponds to the non-axisymmetric case $b=1.2$ while the bottom row (d-f) corresponds to the axisymmetric case $b=1$. The line shows the iso-contour corresponding to 90\% of the local maximum value around the theoretical attractor location which is the origin of the local frame used here.}
    \label{fig:maps}
\end{figure}

A visual inspection of the wave field qualitatively confirms this conclusion.
Figure~\ref{fig:visit} shows the vertical velocity component at a particular time once a periodic response has been obtained.
One can see that the response along the attractor path is much stronger in the $y=0$ plane (see Figure~\ref{fig:visit}(a)) than it is in the $x=0$ plane (see Figure~\ref{fig:visit}(b)).
Note of course that the 2D attractor path (shown as dash lines in Figures~\ref{fig:visit}(a) and (b)) is not the same in both planes since the domain is non-axisymmetric.
This explains why we observe apparently curved wave beams.
They correspond to the gradual modification of the attractor path as the geometry changes along the azimuthal direction due to the weak elliptical deformation.
Similar structures were for example observed by \cite{buhler2007instability}.
We also see in Figure~\ref{fig:visit}(b) that the attractor path in the unstable plane $x=0$ is very close to the bottom corner, which perhaps explains why we have observed a secondary peak of energy in Figure~\ref{fig:dnsprof}.
The bottom corner forces a local shear layer which might overlap with the attractor beam in a non-trivial way.
Nevertheless, Figure~\ref{fig:visit}(c) clearly shows that the amplitude of the stable attractor localised around the plane $y=0$ is significantly larger than that of any other planes, including the unstable attractor localised in the $x=0$ plane.

\begin{figure}
    \centering
    \includegraphics[width=0.75\textwidth]{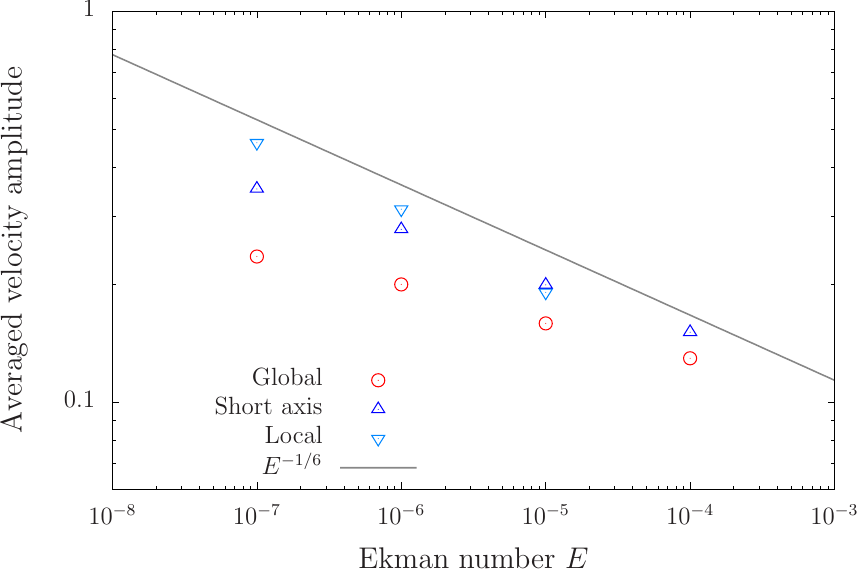}
    \caption{Amplitude scaling with Ekman for three different measures: the average is performed over all attractor paths irrespective of their azimuthal orientation (Global), only along the ray path corresponding to the short axis (Short axis) or focusing on the local maximum observed on the maps shown in Figure~\ref{fig:maps} (Local). Note that the local value at $E=10^{-4}$ is absent since we cannot unambiguously measure a local maximum in that case.}
    \label{fig:scaling}
\end{figure}

In order to show the local structure of the super-attractor, we define a plane perpendicular to the attractor path along the short axis, as shown in Figure~\ref{fig:visit}(a).
The velocity amplitude is averaged over time and normalised by its local maximum closest to the attractor path to obtain the maps displayed in Figure~\ref{fig:maps}.
We compare the axisymmetric case $b=0$ on the bottom row with the non-axisymmetric case $b=1.2$ on the top row.
For the axisymmetric case, one observes the gradual focusing of the axisymmetric wave beam around the attractor position as the Ekman number is decreased.
The observed curvature is due to intersection between the flat plane and the curved axisymmetric wave beam.
This axisymmetric pattern is clearly broken for $b=1.2$.
At $E=10^{-5}$, while we observe a local maximum close to the theoretical position of the super-attractor (which corresponds to the origin with our choice of coordinates), no local beam is observed.
At lower Ekman numbers however, a localised beam is observed with a complex anisotropic structure.
It is more elongated along the transverse direction $y$ than along the in-plane coordinate $s$.
This confirms that the energy injected by the global large-scale forcing is eventually focused preferentially onto the super-attractor path localised around the $y=0$ plane.
While it is too early to convincingly discuss possible scalings with the Ekman number, we nevertheless report the amplitude scaling observed in our simulations in Figure~\ref{fig:scaling}.
We consider three different measures of amplitudes.
The first is obtained by averaging over all 2D attractor paths of each meridional section, which we refer to as global.
The second corresponds to the average on the short axis attractor only while the last corresponds to the local amplitude maximum obtained from the maps displayed in Figure~\ref{fig:maps}.
We observe that for all three measures, the amplitude increases as the Ekman number decreases.
The local measure might follow the scaling $E^{-1/6}$, which is expected from classical 2D attractors \citep{He2023} forced by inviscid forcing, although a much larger range of Ekman numbers should be explored to convincingly conclude on this matter.
This is left for future works.

\section{Discussion}

We have shown how a 2D attractor can be localised in 3D into a particular plane by breaking the axisymmetry of the fluid domain.
All ray paths have been observed to converge to a particular 1D curve contained within a specific plane.
A local analysis has allowed us to confirm the exponential decay of the horizontal angle and has been successfully compared with ray tracing computations.
While numerical simulations have shown a localisation of the energy onto the attracting plane, the harmonic response also contains contributions from the corners which might be at the origin of a secondary peak of energy along the long-axis which is unstable according to our ray tracing analysis.
By construction of our particular geometry, the secondary unstable attractor is more and more affected by the corner contributions as $b$ increases, which might explain why we observe a secondary energy increase there.
However, the observed increase of kinetic energy along the stable super-attractor can only be explained by a secondary focusing since its topology within the particular plane $y=0$ remains the same regardless of the elliptical deformation $b$.

While our example has convincingly shown the existence of such super-attractors, generalising this result to other geometries would be valuable.
We have focused on super-critical reflections only since $\alpha>\theta$ in our case while it is known that the focusing properties of the 3D reflection is non-trivially depending on the criticality of the slope \citep{Pillet2019}.
We suspect that similar super-attractors exist in other conical geometries, including those involving sub-critical slopes, such as those used by \cite{Klein2014}, \cite{sibgatullin2017direct} and \cite{Boury2021}, provided that the axisymmetry used up to now is broken.
Other geometries known to support 2D or axisymmetric attractors, such as the spherical shell or the paraboloidal/parabolic-shaped stadium, can probably generate super-attractors once elliptically deformed.
In that respect, much remains to be done to bridge the gap between the local analysis close to the attracting plane and the global properties for which the whole geometry, and not just its asymptotic behaviour close to the attracting plane, matters.

We have considered the particular case of an elliptic deformation in order to break the axisymmetry.
While it has the advantage of being a well-defined perturbation of the axisymmetric reference case, it is certainly not the only way to proceed.
It would be interesting to consider other types of non-axisymmetric geometries like cuboids \citep{wu2023inertial}.
Possible connections with quantum chaos in stadium billiards resulting in scarred patterns could also be explored \citep{kudrolli2001scarred}.
One could also further investigate the required geometrical properties of the fluid container necessary for the emergence of super-attractors.
Finally, it is important to realise that, while the final super-attractor is indeed one-dimensional in the sense that all rays converge towards a one-dimensional parametric curve, this particular attractor is contained inside a plane since it corresponds to the localisation of the otherwise 2D attractor.
An interesting question would be to know whether general 1D curves, not necessarily contained within the same plane, can be super-attractors and if so in which geometries.
This is a tremendous geometrical problem and we hope this preliminary work will motivate future studies in that direction.

\section*{Acknowledgments}

The authors would like to acknowledge Leo Maas for fruitful discussions and relevant comments on an earlier version of the manuscript.
Centre de Calcul Intensif d'Aix-Marseille is acknowledged for granting access to its high-performance computing resources. This work was granted access to the HPC resources of IDRIS under the allocation 2023-A0140407543 made by GENCI. 

\section*{Declaration of Interests}

The authors report no conflict of interest.

\appendix

\section{3D reflection law} 

In this section,  the 3D reflection law is obtained.  
The relation between the incident and reflected angles $\phi_i$ and $\phi_r$ can be obtained by writing down the conditions on the velocity vectors ${\bf V_i}$
and ${\bf V_r}$ during the reflection process. We consider a plane surface inclined by an angle $\alpha$ (between 0 and $\pi/2$) with respect to the horizontal plane and  
defined it using its normal vector ${\bf n}$ oriented towards the fluid. 
The first condition is the condition of non-penetrability which reads
\begin{equation}
({\bf V_r} + {\bf V_i})\cdot {\bf n} =0 .
\end{equation}
The second condition is the conservation of the horizontal velocity perpendicular to the normal which can be written as
\begin{equation} 
({\bf V_r}\times {\bf e_z})\cdot {\bf n} = ({\bf V_i}\times {\bf e_z})\cdot {\bf n} . 
\end{equation} 
They are two other conditions that prescribe the direction of propagation of the incident and reflected beams which should be on the axial cone of angle $\theta$
\begin{equation} 
 \tan^2\theta ||{\bf V_r}\times{\bf e_z} ||^2 =   |{\bf V_r}\cdot {\bf e_z}|^2 ,
\end{equation} 
\begin{equation} 
 \tan^2\theta ||{\bf V_i}\times{\bf e_z} ||^2 =   |{\bf V_i}\cdot {\bf e_z}|^2 .
\end{equation} 

To obtain a general formula, the idea is to express the vectors ${\bf n}$, $\bf V_r$ and $\bf V_i$ in 
the frame $(x',y',z)$ oriented along the steepest descent direction (see figure \ref{fig:reflection}). 

The above equations can  then be written as
\bsea 
&&(V_{rx'} + V_{ix'}) \tan \varphi_{\bf n}  + (V_{rz} + V_{iz}) = 0 ,\\
&&V_{ry'} = V_{iy'}  ,\\
&&\tan^2\theta (V_{rx'}^2 + V_{ry'}^2)  =  V_{rz}^2 ,\\
&&\tan^2\theta (V_{ix'}^2 + V_{iy'}^2)  =  V_{iz}^2 ,
\label{exp:vrefl}
\esea
where $\varphi_{\bf n}$ is the angle (between 0 and $\pi$) of the vector ${\bf n}$ with respect to the vertical axis.
Using (\ref{exp:vrefl}b) to eliminate $V_{ry'}$ and $V_{iy'}$ in the equation obtained by subtracting (\ref{exp:vrefl}d) to (\ref{exp:vrefl}c) gives
\be
\tan^2\theta(V_{rx'}^2 - V_{ix'}^2) = V_{rz}^2 - V_{iz}^2 ~,
\ee
which, after eliminating $v_{ix'}$ using (\ref{exp:vrefl}a), reduces to
\be
(\tan ^2\theta - \tan^2 \varphi_{\bf n}) V_{rx'} = 2 \tan \varphi_{\bf n}  V_{iz} +  (\tan ^2\theta + \tan^2 \varphi_{\bf n}) V_{ix'} ~.
\ee
Using 
\be 
\frac{V_{iz}}{V_{iy'}} = \frac{{\rm cotan} \varphi_i}{ \sin (\phi_i - \phi_{\bf n})} ~,~~ \frac{V_{ix'}}{V_{iy'}}= {\rm cotan}(\phi_i -\phi_{\bf n})  ~,~~\frac{V_{rx'}}{V_{ry'}} ={\rm cotan}(\phi_r-\phi_{\bf n})
\ee
we get an expression that gives the angle $\phi_r -\phi_{\bf n}$:
\be
\tan (\phi_r -\phi_{\bf n}) = \frac{(\tan^2\theta - \tan^2\varphi_{\bf n}) \sin(\phi_i -\phi_{\bf n})}{  (\tan ^2\theta + \tan^2 \varphi_{\bf n} ) \cos(\phi_i -\phi_{\bf n}) + 2 \tan \varphi_{\bf n} {\rm cotan}\varphi_i }.
\ee 

This formula applies to all configurations.
Only the term $\tan\varphi_n {\rm cotan}\varphi_i$ depends on the orientation of the normal ${\bf n}$ and of the incident beam   ${\bf V_i}$ with respect to the vertical. 
As $ \varphi_i = \pi/2 - \theta$ and $\varphi_{\bf n} = \alpha$ (resp.  $ \varphi_i = \pi/2 + \theta$ and $\varphi_{\bf n} = \pi -\alpha$) when these vectors are oriented upwards (resp. downwards),
we get
\be
\tan (\phi_r -\phi_{\bf n}) = \frac{(\tan^2\theta - \tan^2\alpha) \sin(\phi_i -\phi_{\bf n})}{  (\tan ^2\theta + \tan^2 \alpha ) \cos(\phi_i -\phi_{\bf n}) \pm 2 \tan \alpha \tan \theta },
%\label{exp:tanphir}
\ee 
where the sign $-$ is taken when the two vectors ${\bf n}$ and ${\bf V_i}$ are oriented oppositely with respect to the vertical (one of the two vectors is oriented upward and the other downward). 
This formula works for all types of reflections (subcritical or supercritical), and for horizontal and vertical  surfaces as well. For this reason, we think that it is more convenient than
the formulas that have been used in \cite{Rabitti2014} and \cite{Pillet2018}. 

\section{Local behavior for $b=b_{c\pm}$}

For the values $b= b_{c\pm}$, the two eigenvalues are equal (to $\lambda_{c\pm} = \pm \sqrt{K}$) and equations (\ref{exp:C0D0}a,b) giving $C_0$ and $D_0$ break down. 
In that case, the two eigenvectors given by (\ref{exp:psilambdapm}b) are also equal (to $\bPsi_{c\pm}$). 
One should introduce another vector, say $\bE_1=(1,0)^\top$ to decompose the vector $\bPsi_{2p}$.  
Using the fact that when $b=b_{c\pm}$
\be 
\bcM^{2p} \bE_1 =K^{p}  \bE_1  \mp   p K^{p-1/2}  \sqrt{2(1+K)} \bPsi_{c\pm},
\ee
one can easily show that
\be
\bPsi_{2p} = \left(C_{c\pm} K^{p} \mp D_{c}   \,p K^{p-1/2}  \sqrt{2(1+K)}\right)\bPsi_{c\pm} + D_{c} K^{p} \bE_1 
\ee
where 
\be 
C_{c\pm} = \frac{\phi_0 -\pi}{1\pm \sqrt{K}}~,~~  D_{c} = \psi_0 - \sqrt{\frac{2}{1+K}} (\phi_0 -\pi) .
\ee

\bibliographystyle{jfm}
\bibliography{biblio}

\end{document}